\documentclass[aip,apl,reprint,groupedaddress,floatfix]{revtex4-1}
\usepackage{graphicx}
\usepackage{amsmath}
\usepackage{color}
\newcommand{\io}{In$_2$O$_3$}
\newcommand{\fo}{Fe$_2$O$_3$}

\begin{document}


\title{Enhancing electrical conductivity of room temperature deposited Sn-doped In$_2$O$_3$ thin films by hematite seed layers } 



\author{Christian Lohaus}
\author{C\'{e}line Steinert}
\author{Getnet Deyu}
\author{Joachim Br\"otz}
\author{Wolfram Jaegermann}
\author{Andreas Klein}
\affiliation{Technische Universit{\"{a}}t Darmstadt, Institute of Materials Science, 64287 Darmstadt, Germany}


\date{published online 12 April 2018}

\begin{abstract}
Hematite \fo\ seed layers are shown to constitute a pathway to prepare highly conductive transparent tin-doped indium oxide (ITO) thin films by room temperature magnetron sputtering. Conductivities of up to $\sigma = 3300\,{\rm S/cm}$ are observed. The improved conductivity is not restricted to the interface but related to an enhanced crystallization of the films, which proceeds in the rhombohedral phase.      \\ 
\textcolor{blue}{This article has been published in Applied Physics Letters, Volume 12, Article Number 152105, 2018; The published version is accessible at doi: 10.1063/1.5022683}
\end{abstract}


\maketitle 


Transparent conductive oxide (TCO) thin films are employed in a variety of applications such as displays, solar cells and LEDs.\cite{Gin10} For most TCOs, high electrical conductivity requires high temperature deposition, which limits their application with temperature sensitive substrates. Sn-doped In$_2$O$_3$ (ITO), for example, can be prepared with conductivities $\geq 10^4\,{\rm S/cm}$ at temperatures of $400\,^\circ{\rm C}$,\cite{Ohta00b,Taha98} while conductivities for films deposited at room temperature is typically $< 10^3\,{\rm S/cm}$.\cite{Song99, Nun02, Kim00, Hao08, Gas12, Kim16} The lower conductivity of films deposited at room temperature is mostly caused by lower carrier concentrations. This is related to the amorphous nature of such films and the fact that electron generation in amorphous TCOs is not caused by the heterovalent substituent but rather by the off-stoichiometry, which changes the coordination of the indium atoms.\cite{Kha15, Med17}  

In this work it is shown that the conductivity of ITO thin films, deposited at room temperature by magnetron sputtering without further annealing, can be significantly increased by the insertion of a thin hematite \fo\ seed layer. The achieved electrical conductivities are as good as the best values obtained with doped polycrystalline ZnO and SnO$_2$ layers grown at elevated substrate temperature. \cite{Ell12} The hematite layer, even with only a few nanometer thickness where it doesn't affect the optical properties, enhances the crystallinity of the samples. It also stabilizes the metastable rhombohedral phase of \io. Seed layers have been used before to enhance electrical properties of transparent conductive oxides,\cite{Nakao10} but not for lowering substrate temperatures. 
Samples were prepared by RF-magnetron sputtering in the DArmstadt Integrated SYstem for MATerials Research (DAISY-MAT), which combines a multitechnique surface analysis system with multiple thin film deposition chambers.\cite{Kl13} In a first step ultrathin hematite films with $\leq 10\,{\rm nm}$ thickness were deposited from a metallic iron target in an Ar-O$_2$ mix with $8\,\%$ oxygen at a pressure of $0.5\,{\rm Pa}$. Amorphous Quartz and c-cut Sapphire (S0001) were used as substrates. If not noted otherwise, the deposition temperature was $400\,{\rm ^\circ C}$. With these parameters phase pure, crystalline hematite ($\alpha$-\fo) thin films are formed.\cite{Lohaus2017} After cooling to room temperature the samples were transferred in ultra-high vacuum (UHV) to a second deposition chamber for ITO deposition. This was performed using a ceramic ITO target with $10\,{\rm wt.\%}$ SnO$_2$ doping. The depositions were performed at room temperature in pure Argon atmosphere at a pressure of $0.5\,{\rm Pa}$ and a power density of $1.23\,{\rm W/cm^2}$.

X-Ray and ultraviolet photoelectron spectroscopy (XPS, UPS) were performed without breaking vacuum using a Physical Electronics PHI 5700 spectrometer. Core level and XP valence band spectra were recorded using monochromatized Al K$_{\alpha}$ radiation. UP valence band spectra were acquired by using He~I radiation from a Helium discharge lamp. Calibration of the binding energies was achieved by measuring the Ag~3d$_{5/2}$ core level and the Fermi edge of a sputter cleaned Ag foil at the same day of the measurements. X-ray diffraction was carried out in $\theta-2\theta$ geometry using a Seifert XRD 30003 PTS-S with Cu K$_\alpha$ radiation. Conductivity and Hall-measurements were performed on a home made setup in van-der-Pauw geometry.\cite{Hoh14} The sample current was $1\,{\rm mA}$ in all measurements. Optical measurements were performed using a Agilent Cary 7000 Universal Measurement Spectrophotometer.

\begin{figure*}[ht]
\centering
\includegraphics[width=15cm]{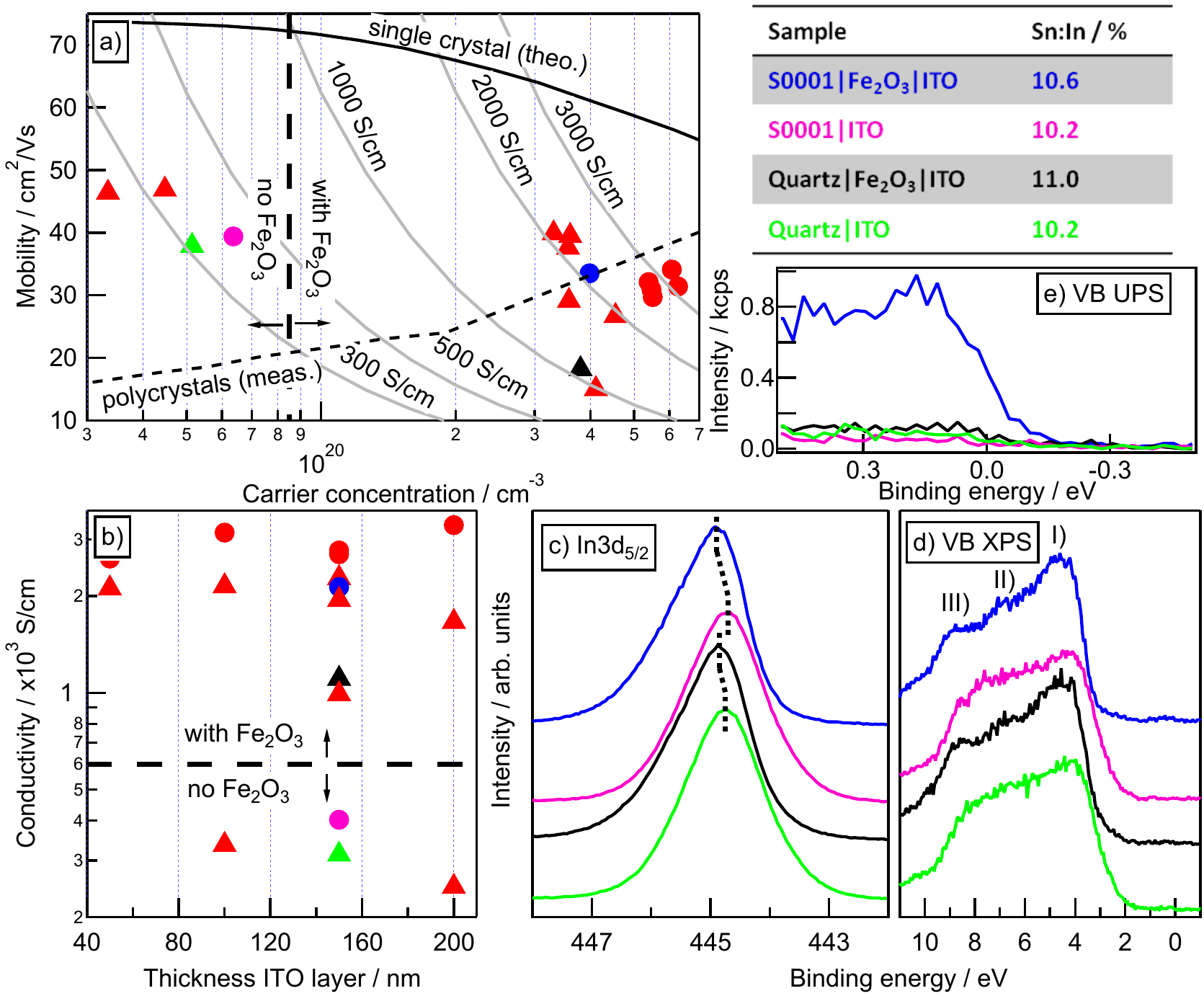}%
\caption{Electrical measurements (a, b), In~3d$_{5/2}$ core levels (c), XP valence bands (d), and magnified UP Fermi edge spectra (e) of room temperature deposited ITO thin films with and without \fo\ seed layers. In (a) and (b) films grown on sapphire (0001) are indicated by circles and films grown on quartz by triangles. The colored symbols in (a) and (b) and the spectra in (c)-(e) correspond to the same samples as identified in the table giving the Sn:In ratio determined from XP core level intensities. \label{fig:xps_cond}}
\end{figure*}

Electrical as well as surface chemical and electronic properties of ITO thin films with and without \fo\ seed layer are presented in Figure \ref{fig:xps_cond}. The conductivity of the films with seed layer are about an order of magnitude higher compared to those deposited on the bare quartz and sapphire substrates. A higher carrier concentration is the origin of the increased conductivity. The carrier mobility is slightly reduced from $40-46\,{\rm cm^2/Vs}$ $29-40\,{\rm cm^2/Vs}$. The conductivity of ITO films grown on \fo\ seed layers is rather independent on film thickness as evident from Fig.~\ref{fig:xps_cond}(b). The enhanced conductivity is therefore clearly a bulk phenomena and not related to the region next to the \fo\ interface.

XP spectra, in which the effect has actually been identified originally, confirm the high conductivity. Films grown on \fo\ seed layers show consistently higher core level and valence band maximum binding energies compared to their counterparts without \fo\ seed layer. The higher conductivity of the spectra is also evident from the stronger asymmetry of the core level lines towards higher binding energy,  which relates to a higher plasmon energy and therefore carrier concentration.\cite{Kor10} Finally,  occupied conduction band states are also observed by UPS as a Fermi edge emission shown in Fig.~\ref{fig:xps_cond}(e). The sample grown on sapphire (0001) with \fo\ layer shows the highest intensity of the Fermi edge, corresponding well with the highest charge carrier concentration. 

\begin{figure}
\centering
\includegraphics[width=8cm]{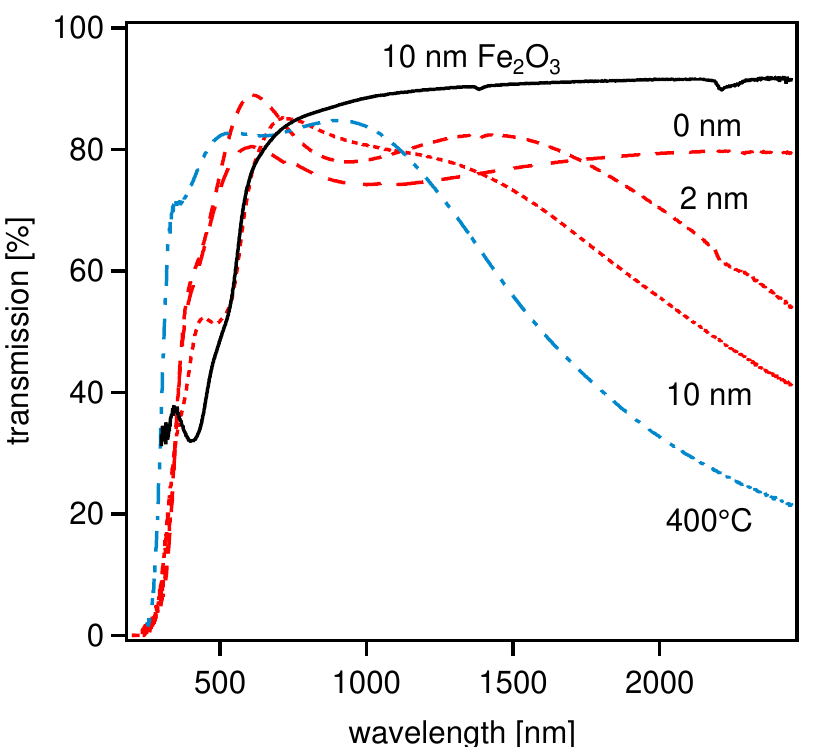}%
\caption{Optical transmission spectra of $150\,{\rm nm}$ thick ITO films grown at room temperature on sapphire with $0$, $2$ and $10\,{\rm nm}$ thick \fo\ seed layers. Transmission spectra recorded for a $10\,{\rm nm}$ thick \fo\ film on sapphire and of a $150\,{\rm nm}$ thick ITO film grown at $400\,{\rm ^\circ C}$ on quartz without seed layer are added for comparison.}
\label{fig:optic}
\end{figure}

Optical transmission spectra of different films are compared in Fig.~\ref{fig:optic}. Relative to the ITO film grown at room temperature on sapphire without seed layer, an even improved transmission is observed for a seed layer thickness of $2\,{\rm nm}$. Increasing the seed layer thickness to $10\,{\rm nm}$ reduces the transmission for wavelengths between $370-630\,{\rm nm}$. In this range, the transmission follows the one recorded for a $10\,{\rm nm}$ thick \fo\ layer, indicating that the transmission of the ITO layer is still high and that an optimized \fo/ITO stack can fully satisfy the requirements of a transparent electrode. 

The Sn/In ratio determined from XPS for films with and without \fo\ seed layer on quartz and sapphire are all within $10-11\,{\rm at.\%}$ as indicated in the table included in Fig.~\ref{fig:xps_cond}. This corresponds well with the nominal target composition. In contrast, films with high conductivity grown at higher substrate temperatures, exhibit a significantly enhanced Sn/In ratio, indicating a segregation of Sn to the surface.\cite{Gas06, Fri16} Such a segregation is evidently not observed when the films are grown at room temperature. This is reasonable given the fact that Sn-donors are mobile only at $T \gtrsim 300\,{\rm ^\circ C}$.\cite{Gas06,Hoh14} 

\begin{figure}[htp]
\centering
\includegraphics[width=7.5cm]{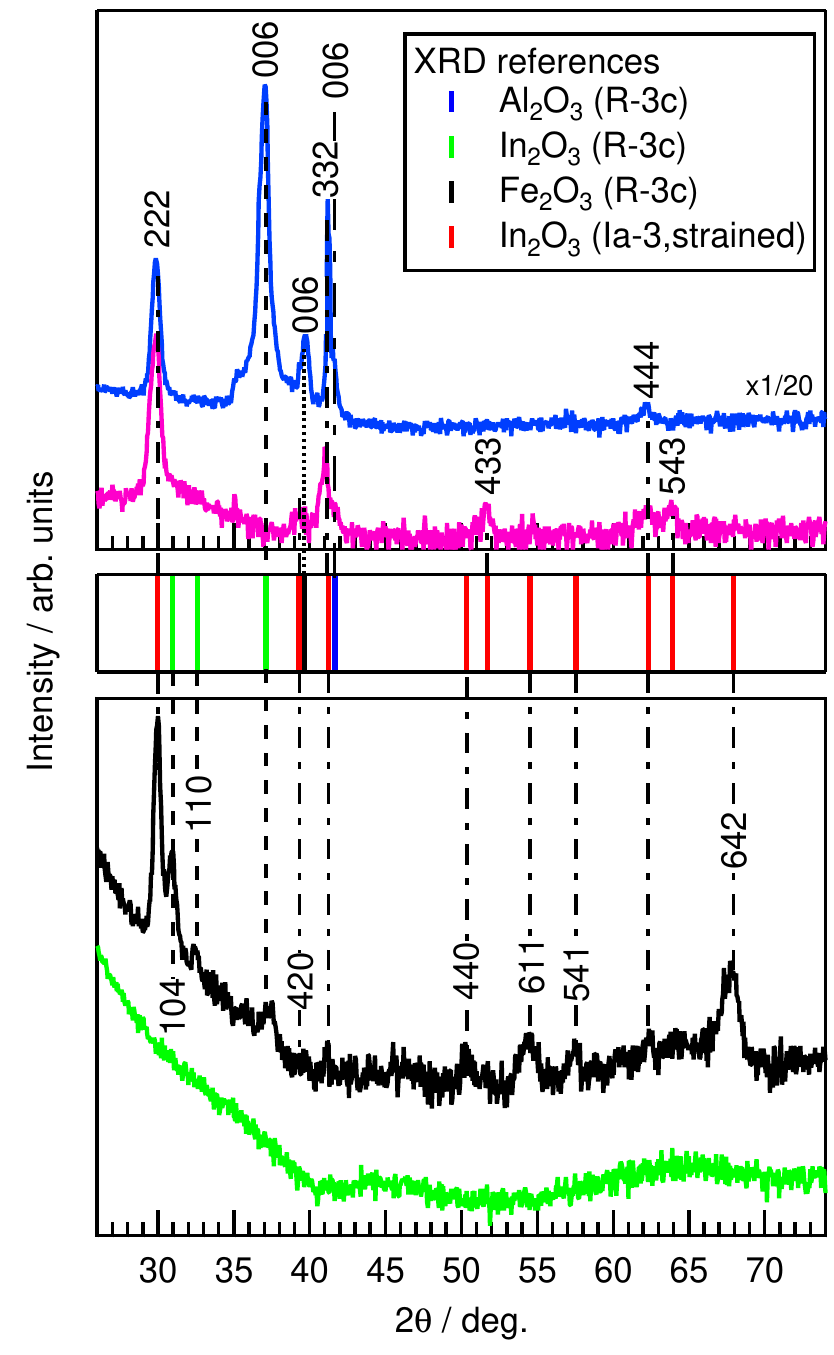}
\caption{XRD pattern from $\protect\theta-2\theta$-measurements on samples with a ITO thickness of $150\,{\rm nm}$. From bottom to top the samples are the following: quartz/ITO (green), quartz/\fo/ITO (black), S0001/ITO (magenta), and S0001/\fo/ITO (blue). All \fo\ layers had a thickness of $10\,{\rm nm}$. All reflections have been identified and labeled with the help of pdf-card reference data (pdf card numbers rh-Al$_{2}$O$_{3}$: 11-661, rh-\fo: 84-311, rh-\io: 73-1809, and cu-\io: 6-416 (strained)).\label{fig:XRD}}
\end{figure}

The XP valence band spectra of samples grown with and without \fo\ seed layer exhibit a different shape. In particular, three distinct features are observed for films grown on \fo, which are less pronounced in the spectra of samples without \fo\ layer. The differences might be related to different crystallinity, orientation or phase composition. In order to reveal the influence of the hematite layer on the structural properties of the ITO, X-ray diffraction has been performed on $150\,{\rm nm}$ thick films. The corresponding diffraction patterns are shown in Figure \ref{fig:XRD}.

The diffraction pattern for the ITO film on quartz without the iron oxide layer does not show any reflections, indicating an amorphous structure. In contrast to this, a number of reflections appear after insertion of the \fo\ layer. These can be assigned to cubic bixbyite \io\ with $222$ being the most intense one. There are, however, also some reflections, which do not belong to this phase. These additional signals can be assigned to the rhombohedral \io\ phase. 

In contrast to the ITO film deposited on quartz, the sample on S0001 without \fo\ layer shows some reflections. Besides a very small reflection from the Sapphire substrate, only signals from the bixbyite \io\ phase are observed. Again, $222$ from this phase is the most intense reflection. The addition of a \fo\ layer changes the diffraction pattern considerably. The intensity of all reflections is increased by a factor of 20. Most intense is now the  (006) reflection of the rhombohedral ITO phase. In addition, the rhombohedral $006$ reflections of \fo\ and Al$_2$O$_3$ can be identified. This indicates epitaxial growth of rhombohedral \io\ and \fo\ on sapphire (0001), which all share the same space group ($\mbox{R}\overline{3}\mbox{c}$).\cite{DeBoer2016,Kim1997c} Among the reflections of the ITO bixbyite phase, the $332$ is the most intense one. 

The diffraction angles of rhombohedral \io\ match those of the powder reference (pdf-card 73-1809). In contrast, the reflections from the cubic phase are shifted to smaller angles, which corresponds to an expansion of the lattice parameter by about $2\,\%$. This is observed for all samples showing reflections from bixbyite \io.

The insertion of an ultrathin crystalline \fo\ film significantly enhances the crystallinity of room temperature deposited ITO films. However, while the cubic bixbyite structure is commonly observed when films are grown at elevated temperature, the \fo\ seed layer favors the growth of rhombohedral ITO, which is a metastable modification.\cite{Gurlo2008} The different shape of the valence band spectra is also consistent with the presence of rhombohedral ITO.\cite{King09}

Only few studies report on the electric properties of rhombohedral ITO thin films. Solution processed nanoparticles fired at $600\,{\rm ^\circ C}$ and aged at $500\,{\rm ^\circ C}$ show a mixture of cubic and rhombohedral phases with a slightly smaller resistance but higher sensitivity towards gaseous ethanol compared to purely cubic films.\cite{Huh2001,Kim2003b,Seo2006} A transition from cubic to rhombohedral structure with film thickness has been reported by Malik and coworkers.\cite{Malik1997}. They could show that the mobility of thicker rhombohedral films ($37\,{\rm cm^2/Vs}$) was higher than for thinner cubic films ($19\,{\rm cm^2/Vs}$). The charge carrier concentration was slightly lower for the rhombohedral films ($7.7\cdot 10^{20}\,{\rm cm^{-3}}$) compared to the cubic ones ($9\cdot 10^{20}\,{\rm cm^{-3}}$). Dunlop and coworkers have reported partially rhombohedral thin films at temperatures of $415 - 500 \, {\rm ^\circ C}$, showing a carrier mobility of $27\,{\rm cm^2/Vs}$ and a carrier concentration of $5.7\cdot 10^{20}\,{\rm cm^{-3}}$. \cite{Dunlop2010} Both values are higher compared to purely cubic reference samples, which was attributed to lower internal strain. The highest conductivities reported for at least partially rhombohedral ITO is therefore in the range of $2400-4500 \, {\rm S/cm}$. These values are comparable to those reported here. However, the previous values have all been obtained with processing temperatures well above $400\,{\rm ^\circ C}$, while the conductivity achieved in the present work was achieved with room temperature deposited ITO without further annealing. 

Based on the studies reported here, different mechanisms might contribute to the enhanced conductivity induced by the \fo\ seed layer: (i) the enhanced crystallinity, (ii) the absence of dopant segregation, (iii) the presence of the rhombohedral phase. The literature survey given above provides no evidence for fundamentally superior electrical properties of the rhombohedral compared to the cubic bixbyite \io\ phase. It is therefore believed that the rhombohedral structure is not responsible for the higher conductivity. 

Carrier mobilities in amorphous TCOs are comparable to those of their crystalline counterparts.\cite{Nom04} This is caused by the low concentration of tail states, which is a consequence of the s-orbital character of the conduction band states.\cite{Nom07,Kami10,Med17} On the first glance, it is therefore not surprising that the carrier mobility is not much different for films grown with and without \fo\ seed layer. However, crystallization induces grain boundaries, which act as carrier traps and scattering centers.\cite{Seto75} The carrier mobilities of our films are noticeably lower than those of single crystalline material (see solid line in Fig.~\ref{fig:xps_cond}(a) and Ref.~\cite{Prei13}). Grain boundary scattering is therefore active. There is evidence that the segregation of the Sn dopants to grain boundaries reduces carrier maobility.\cite{Hoh14,Fri16} But the segregation requires elevated temperatures to activate Sn diffusion. No Sn segregation is therefore expected, which agrees with the absence of surface segregation of Sn. In any case, the main improvement in electrical conductivity induced by the \fo\ layer is the higher carrier concentration. Carrier mobility and the absence of segregation should therefore also play a minor role. 

Films deposited at room temperature are normally amorphous. In such films, the free carriers are not induced by heterovalent dopants, such as Sn, but by the off-stoichiometry (oxygen deficiency) of the samples.\cite{Kha15, Med17} Crystallization will activate the donors. It is therefore straightforward to assign the origin of the higher carrier concentration of films grown on \fo\ seed layer to the enhanced crystallinity. Given the significantly higher diffraction intensities, the enhanced crystallinity of films grown on sapphire/\fo, compared to those grown on quartz/\fo, also explains the higher carrier concentration of the former.   

Other seed layers may exits, which favor crystallization at room temperature in the bixbyite phase. These might result in similar improvement. There are indeed literature reports on ITO films prepared at room temperature, which exhibit similar conductivity than the films prepared here with \fo\ seed layer (see Refs. \cite{Taha98, Wu99} and references therein). The crystalline state of these films is not reported, however. Our deposition conditions were optimized for deposition at elevated temperatures. It would, therefore, be compelling to see whether there is room for additional improvement by a seed layer in a room temperature optimized deposition.   

In summary, we have shown that the use of a \fo\ seed layer can enhance the electrical conductivity of ITO thin films deposited at room temperature by about one order of magnitude reaching yet a maximum conductivity of $3300\,{\rm S/cm}$. The hematite structure of the \fo\ seed layer induces a local epitaxial growth and thereby enhances the crystallinity of the \io\ layer. The effect is most pronounced on c-cut sapphire substrates but also observed on amorphous quartz substrates. The \fo\ layer results in a preferred formation of the metastable rhombohedral \io\ phase, which is not considered to affect the electrical properties. However, crystallization at low temperature offers the advantage of suppressing dopant diffusion and therefore segregation, which can result in higher carrier mobilities compared to films crystallized at higher temperatures. The use of seed layers therefore offers an approach to reach high conductivity low temperature processed TCO films. The possibility to stabilize the rhombohedral $\mbox{R}\overline{3}\mbox{c}$ phase of \io\ by the \fo\ seed layer also provides a simple route for the preparation of such films.     


This work was supported by the European Commission within the Knowledge Innovation Center (KIC) raw matter. 

\bibliography{references}

\end{document}